\documentclass[english,pra,twocolumn,superscriptaddress]{revtex4-1}
\usepackage{amsmath}
\usepackage{amssymb}
\usepackage{graphicx}
\usepackage{esint}
\usepackage{times}

\begin{document}

\title{Dynamic entanglement transfer in a double-cavity optomechanical system}

\author{Tiantian Huan}

\affiliation{College of Information Engineering, East China JiaoTong University,
Nanchang, China}

\affiliation{Institute of Applied Physics and Materials Engineering, FST, University
of Macau, Macau}

\author{Rigui Zhou}

\thanks{Corresponding author}

\affiliation{College of Information Engineering, East China JiaoTong University,
Nanchang, China}

\author{Hou Ian}

\affiliation{Institute of Applied Physics and Materials Engineering, FST, University
of Macau, Macau}
\begin{abstract}
We give a theoretical study of a double-cavity system in which a mechanical
resonator beam is coupled to two cavity modes on both sides through
radiation pressures. The indirect coupling between the cavities via
the resonator sets up a correlation in the optomechanical entanglements
between the two cavities with the common resonator. This correlation
initiates an entanglement transfer from the intracavity photon-phonon
entanglements to an intercavity photon-photon entanglement. Using
numerical solutions, we show two distinct regimes of the optomechanical
system, in which the indirect entanglement either builds up and eventually
saturates or undergoes a death-and-revival cycle, after a time lapse
for initiating the cooperative motion of the left and right cavity
modes.
\end{abstract}


\maketitle

\section{Introduction}

Cavity optomechanical systems~\cite{kippenberg08} arise from the
classical Fabry-Perot interferometer~\cite{vaughan89} by replacing
one of the fixed sidewalls with a cantilever or double-clamped beam~\cite{cleland96,meyer88,kleckner06}.
The one-dimensional degree of freedom introduced by the movable mechanical
element adds a free resonator mode to the cavity system and allows
this mode to interact with the cavity field through radiation pressure
on the reflectively coated mechanical resonator. Regarded as a micromirror,
this resonator can be feedback-controlled through the cavity field,
on which numerous cooling protocols have been conceived and experimentally
demonstrated in the last decade~\cite{metzger04,naik06,arcizet06,ian08-1,liu13}.

The degree of control in this hybrid cavity-micromirror system can
be further enhanced when the micromirror is replaced by a double-face
reflective membrane~\cite{thompson08,jayich08}. If a second optical
cavity is coupled to it on the opposite side of the existing cavity,
a two-mode or double-cavity optomechanical system with enhanced nonlinearity
is formed~\cite{pinard05,miao09,naeini11,ludwig12}. Entanglement-wise,
though it was observed that the enhanced squeezing resulted from the
nonlinear coupling helps generate static entangled state of distant
mirrors\cite{pinard05}, the dynamic property of entanglement between
the two cavities is less well-understood.

Recent studies reveal that the dynamics of phonon-photon entanglement
plays an important role in defining the system characteristics, such
as the transitions between oscillation modes~\cite{wang14,ying14},
robustness against noisy environment~\cite{tian13}, sudden death
and revival of states~\cite{ian08,chang09,lin14}, and optimal entanglement~\cite{ydwang13}.
In this article, we study the dynamics of the entanglements in a double-cavity
optomechanical system where each photon mode in the two opposite cavities
is, structure-wise, symmetrically coupled to a common mechanical resonator
mode via radiation pressures, albeit asymmetric coupling strengths
and driving powers are generally assumed. Our main concern is to determine
how the cavity-resonator entanglements~\cite{vitali07} can be transferred
to the indirectly coupled cavities over time.

We show here such an entanglement transfer is possible in a double-cavity
optomechanical system through measuring the entanglements in logarithmic
negativities among the component pairs. In particular, the negativity
is computed through determining the symplectic eigenvalues of a covariance
matrix that relates the fluctuations of all six quadratures of the
system's main components. This method is standard in the literature
of dynamic entanglement but we have generalized it to apply on a $6\times6$
covariance matrix. We observe that the successful generation of entanglement
transfer only requires a single-sided driving laser and that the transfer
patterns can be distinctively categorized into two groups for the
different operating regimes assumed by optomechanical system. 

Moreover, all the logarithmic negativities computed exhibit a time
delay before the first appearance of a non-zero value. This time point
signifies the initiation of cooperative motions among the three components
in the optomechanical system, showing the transient response of the
system to the external driving lasers as a whole. Nonetheless, the
indirect entanglement between the left and the right cavity is apparent
in all cases, thereby facilitating a mechanism for entanglement relay
through cascaded cavities although the cavities are physically not
directly coupled. Such a mechanism would be useful to quantum information
processing, especially in terms of non-adiabatic quantum state transfer~\cite{palomaki13,zhang03},
and would provide a physical means to realize cavity arrays or resonator
waveguides for transmitting information encoded in a quantum state~\cite{zhou08,gong08,xuereb12}.

In Sec.~\ref{sec:model}, we give a detailed description of the double-cavity
model. The equations of motions are derived under the Heisenberg picture
in Sec.~\ref{sec:dynamics} and the steady-state solutions are calculated
to give proof of the sufficiency of single-sided driving. After the
covariance matrix of the fluctuations is introduced, the entanglements
among all component pairs are computed numerically and analyzed in
Sec.~\ref{sec:entanglement_transfer}. The conclusions are given
finally in Sec.~\ref{sec:conclusions}.

\section{Double optomechanical cavity\label{sec:model}}

\begin{figure}
\includegraphics[bb=0bp 70bp 635bp 305bp,clip,width=8.5cm]{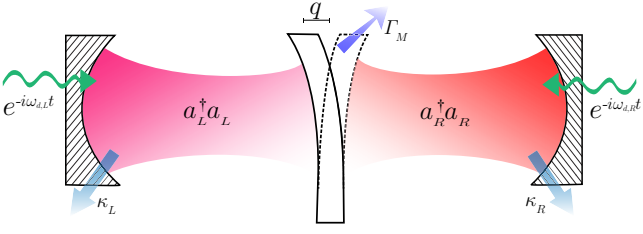}

\protect\caption{(Color online) Model schematic of the double-cavity optomechanical
system: a mechanical element with reflective coatings on both sides
serves as a double-face mirror that experiences radiation pressures
from both the left-side cavity and the right-side cavity. An incident
driving laser enters the double-cavity system from the left side.~\label{fig:model}}
\end{figure}

The proposed double-cavity optomechanical system is illustrated in
Fig.~\ref{fig:model}, in which a mechanical resonator with reflective
coatings on both sides receives the radiation pressures from both
the cavity on the left side (L) and the cavity on the right side (R).
The total Hamiltonian $H=H_{0}+H_{\mathrm{rad}}+H_{\mathrm{ext}}$
thus consists of three parts, which reads ($\hbar=1$), respectively,
\begin{eqnarray}
H_{0} & = & \omega_{L}a_{L}^{\dagger}a_{L}+\omega_{R}a_{R}^{\dagger}a_{R}+\frac{p^{2}}{2m}+\frac{1}{2}m\Omega_{M}^{2}q^{2},\label{eq:Ham_0}\\
H_{\mathrm{rad}} & = & \left(\eta_{L}a_{L}^{\dagger}a_{L}-\eta_{R}a_{R}^{\dagger}a_{R}\right)q,\label{eq:Ham_rad}\\
H_{\mathrm{ext}} & = & i\varepsilon_{L}\left(a_{L}^{\dagger}e^{-i\omega_{d,L}t}-\mathrm{h.c.}\right)+i\varepsilon_{R}\nonumber \\
 &  & \times\left(a_{R}^{\dagger}e^{-i\omega_{d,R}t}-\mathrm{h.c.}\right)\label{eq:Ham_ext}
\end{eqnarray}

The part $H_{0}$ accounts for the free Hamiltonians of the resonator
and the cavities, the latter being regarded as bosonic modes of frequencies
$\omega_{L}$ and $\omega_{R}$. We associate a pair of annihilation
and creation operators $a_{\sigma}$ and $a_{\sigma}^{\dagger}$ for
each bosonic mode, where $\sigma$ indexes the cavity side, either
left $L$ or right $R$. We assume the frequencies $\omega_{L}$ and
$\omega_{R}$ to be different in general according to the asymmetric
cavity lengths $\ell_{\sigma}$ and finesses $F_{\sigma}$ assumed.
The leakage rates are defined correspondingly from these parameters:
$\kappa_{\sigma}=\pi c/2F_{\sigma}\ell_{\sigma}$. The mechanical
resonator is described by the conjugate pair $q$ and $p$, along
with its oscillation mode frequency of $\text{\ensuremath{\Omega}}_{M}$
and its mechanical damping rate of $\Gamma_{M}$.

The part $H_{\mathrm{rad}}$ accounts for the phonon-photon interactions
derived from the radiation pressures. The radiation pressure from
each cavity side results from the deformation of the cavity volume
due to the displacement $q$ of the middle resonator, which shifts
the resonance frequencies of each cavity modes. Expending this frequency
shift to first order, i.e. $g_{L}\simeq\omega_{L}/\ell_{L}$, the
radiation pressure term sets the optomechanical coupling on the left
side with strength $\eta_{L}=g_{L}/\sqrt{2m\Omega_{M}}$, where $m$
is the effective mass of the resonator mode. The same derivation applies
to the right cavity, giving the coupling strength $\eta_{R}=g_{R}/\sqrt{2m\Omega_{M}}$
but the leading sign would be opposite to $\eta_{L}$ as the common
resonator has its displacement $q$ follow opposite directions for
the two radiation pressures. Between the two cavities, there is no
direct coupling.

The part $H_{\mathrm{ext}}$ accounts for the two external driving
lasers with frequency $\omega_{d,L}$ and frequency $\omega_{d,R}$.
The driving strength $\varepsilon_{\sigma}$ of each laser is related
to the input laser power $P_{\sigma}$ and the leakage $\kappa_{\sigma}$
by $|\varepsilon_{\sigma}|^{2}=2\kappa_{\sigma}P_{\sigma}/\hbar\omega_{d,\sigma}$.
Note that we assume an asymmetric setting for the double-cavity system:
the radiation pressures from the two sides are not identical and the
cavities are unequally driven.

To study the indirect entanglement across the two cavity modes, we
begin with the dynamics of the three components in the double-cavity
optomechanical system through deriving a set of nonlinear Langevin
equations. We carry out this step by finding the Heisenberg equations
of the operators from the Hamiltonian in E.~(\ref{eq:Ham_0})-(\ref{eq:Ham_ext})
and introducing phenomenologically the relaxation terms and their
associative Brownian noise terms. The Langevin equations under the
rotating frames of reference read 
\begin{eqnarray}
\overset{.}{q} & = & \frac{p}{m},\nonumber \\
\overset{.}{p} & = & -m\Omega_{M}^{2}q-\Gamma_{M}p-\eta_{L}a_{L}^{\dagger}a_{L}+\eta_{R}a_{R}^{\dagger}a_{R}+\xi,\nonumber \\
\overset{.}{a_{L}} & = & -(\kappa_{L}+i\Delta_{L})a_{L}-i\eta_{L}a_{L}q+\varepsilon_{L}+\sqrt{2\kappa_{L}}a_{L}^{\mathrm{in}},\nonumber \\
\overset{.}{a_{R}} & = & -(\kappa_{R}+i\Delta_{R})a_{R}+i\eta_{R}a_{R}q+\varepsilon_{R}+\sqrt{2\kappa_{R}}a_{R}^{\mathrm{in}},\label{eq:Langevin}
\end{eqnarray}
where $\Delta_{0,L}=\omega_{L}-\omega_{d,L}$ ($\Delta_{0,R}=\omega_{R}-\omega_{d,R}$)
is the static detuning of the left (right) cavity field from the left
(right) driving laser. The zero-mean fluctuation terms $a_{\sigma}^{\mathrm{in}}$
obey the correlation relation $\langle a_{\sigma}^{\mathrm{in}}(t)a_{\sigma}^{\mathrm{in}\dagger}(t')\rangle=\delta(t-t').$

The mechanical mode is under the influence of stochastic Brownian
noise that satisfies in general the non-Markovian auto-correlation
relation with a colored spectrum: 
\begin{equation}
\left\langle \xi(t)\xi(t')\right\rangle =\frac{\Gamma_{M}}{\Omega_{M}}\int d\omega\frac{\omega}{2\pi}e^{-i\omega(t-t')}\left\{ \coth\left(\frac{\omega}{2k_{B}T}\right)+1\right\} ,
\end{equation}
where $k_{B}$ is the Boltzmann constant and $T$ is the temperature
of the mechanical bath. However, for a high quality mechanical resonator
with $Q=\Omega_{M}/\Gamma_{M}\gg1$, this non-Markovian process can
be approximated as a Markovian one, where its fluctuation-dissipation
relation can be asymptotically simplified to~\cite{Giovannetti01,Fabre94}:
\begin{equation}
\left\langle \xi(t)\xi(t')+\xi(t')\xi(t)\right\rangle /2=\Gamma_{M}(2\bar{n}+1)\delta(t-t'),\label{eq:correlation}
\end{equation}
where $\bar{n}=\left(\exp\{\Omega_{M}/k_{B}T\}-1\right)^{-1}$ is
the mean occupation number of the mechanical mode. This simplified
Markovian relation will be assumed in the calculation of the entanglements.

\section{Dynamics and entanglement\label{sec:dynamics}}

\subsection{Steady states}

In a single-cavity optomechanical system, the radiation pressure contributes
the nonlinear photon number term in the Langevin equation of the mirror
momentum $p$ in Eq.~(\ref{eq:Langevin}), leading to a multistability
of the coordinate $p$ with three nonzero steady states. For a double-cavity
case here, the second radiation pressure by the other cavity contributes
a similar term in the equation. Under the asymmetric setting, the
two pressure terms are not commensurate and the number of steady states
of $p$ increases to five. The steady states are given by
\begin{eqnarray}
\langle q\rangle & = & \frac{-\eta_{L}\left|\left\langle a_{L}\right\rangle \right|^{2}+\eta_{R}\left|\left\langle a_{R}\right\rangle \right|^{2}}{m\Omega_{M}^{2}},\label{eq:steady_q}\\
\langle a_{\sigma}\rangle & = & \frac{\varepsilon_{\sigma}}{\kappa_{\sigma}+i(\Delta_{0,\sigma}\pm\eta_{\sigma}\left\langle q\right\rangle )},\label{eq:steady_a}
\end{eqnarray}
where the plus (minus) signs in the second equation refers to the
left (right) cavity.

For entanglement generation, it is necessary for the equation set
(\ref{eq:steady_q})-(\ref{eq:steady_a}) to have non-zero steady
states. Therefore, a single-cavity optomechanical system usually requires
an external driving laser (i.e., non-zero value of $\varepsilon$)
to drive the mechanical resonator out of its zero steady states at
equilibrium position. However, for optomechanical systems with double-sided
cavities, one external driving laser at either end of the cavities
is sufficient to drive the mechanical resonator out of its zero position,
in which case Eq.~(\ref{eq:steady_q}) would fall back to the single-cavity
case of three roots.

In addition, we can observe that even when the double cavities have
exactly symmetrical setup, i.e. identical laser driving amplitudes
($\varepsilon_{L}=\varepsilon_{R}=\varepsilon$), radiation pressures
$(\eta_{L}=\eta_{R}=\eta)$, and cavity lengths, the differing signs
before $\eta_{\sigma}\left\langle q\right\rangle $ to be taken by
$\left\langle a_{L}\right\rangle $ and $\left\langle a_{R}\right\rangle $
in Eq.~(\ref{eq:steady_a}) allows the cavities to admit non-zero
steady states. This is because the two cavity modes are constructively
interfering with each other at the interface of the mechanical resonator
through their indirect interactions of radiation pressures. In other
words, even though the radiation pressures are exerted along opposite
directions, the dynamic $\pi$-phase difference between the cavities
fields, reflected in the Hamiltonian Eq.~(\ref{eq:Ham_rad}) as the
generator of the cavity motion, render the radiation pressures out
of phase to favor the generation of entanglement. Given the symmetric
setting where $\kappa_{L}=\kappa_{R}=\kappa$ and $\Delta_{0,L}=\Delta_{0,R}=\Delta_{0}$
in addition to the identities in driving amplitudes and radiation
pressures, the condition for the steady-state equations to admit real
roots is the inequality among the system parameters
\begin{equation}
\eta\varepsilon\geq\sqrt{\frac{m\Omega_{M}^{2}}{4\Delta_{0}}}(\kappa^{2}+\Delta_{0}^{2}).
\end{equation}
Its derivation is given in Appendix A. Finding $m\Omega_{M}^{2}$
as the Young's modulus of the resonator ($m\Omega_{M}^{2}<\Delta_{0}$)
and that the cavities have sufficient finesses ($\kappa\leq\Delta_{0}$),
the above criterion is met in most scenarios and the validity of entanglement
generation is almost guaranteed.

For the symmetric setting, we expect the patterns of entanglement
generations between either end of the cavity modes and the mechanical
resonator to be qualitatively similar and differ only quantitatively
in their variations over time. Deviating from this setting, the increase
in asymmetry among the system parameters would increase the qualitative
difference between the patterns of entanglements. We demonstrate these
effect later in Sec.~\ref{sec:entanglement_transfer}.

\subsection{Entanglement measure}

Theoretically, the entanglements in terms of logarithmic negativity
are computed through the fluctuations of the cavity quadratures about
the steady states obtained from Eqs.~(\ref{eq:steady_q})-(\ref{eq:steady_a}).
That is, we define the dimensionless quadratures of the two cavity
fields as
\begin{eqnarray}
X_{\sigma} & = & \frac{1}{\sqrt{2}}\left(a_{\sigma}+a_{\sigma}^{\dagger}\right),\\
Y_{\sigma} & = & \frac{1}{i\sqrt{2}}(a_{\sigma}-a_{\sigma}^{\dagger}).
\end{eqnarray}
and the corresponding input noise operators accordingly. Then taking
$\mathcal{O}\equiv(q,p,X_{L},Y_{L},X_{R},Y_{R})$ as the vector operator
for all the quadratures in the optomechanical system, we expand it
to first-order using a c-number steady-state value and a zero-mean
fluctuation operator $\mathcal{O}(t)=\left\langle \mathcal{O}\right\rangle +\delta O(t)$.
In addition, the nonlinear terms are linearized assuming $|\left\langle a\right\rangle |\gg1$
in the expansion: $\left\langle a^{\dagger}a\right\rangle \simeq|\left\langle a\right\rangle |^{2}$
and $\left\langle aq\right\rangle \simeq\left\langle a\right\rangle \left\langle q\right\rangle $,
while the higher-order products of the fluctuations are ignored. 

The Langevin equations in Eq.~(\ref{eq:Langevin}) with the first-order
expansion gives a coupled system of differential equations about the
noise operators, enabling the coupling between the fluctuations of
the two cavity fields and the mechanical resonator and thus the generation
of entanglement between the two optical modes. Note that even though
we have linearized the equations for these operators, eliminating
the mechanical quadratures $q$ and $p$ in Eq.~(\ref{eq:Langevin})
will lead to equations of $a_{L}$ and $a_{L}^{\dagger}$ nonlinearly
related to $a_{R}$ and $a_{R}^{\dagger}$. This implies that the
indirect entanglement between the quadratures of the left and the
right cavities follow a nonlinear form in time.

In the following, instead of solving the coupled equations analytically,
we follow the standard numerical approach adopted by the current researches
on dynamic entanglement~\cite{wang14,ying14,vitali07}. The difference
here is that we have a 6-component vector $u=(\delta q,\delta p,\delta X_{L},\delta Y_{L},\delta X_{R},\delta Y_{R})$
over the six quadratures of the tripartite optomechanical system instead
of the usual 4-component vector. Similarly extending the input-noise
vector to the 6-component $n=(0,\xi,\sqrt{2\kappa_{L}}X_{L}^{\mathrm{in}},\sqrt{2\kappa_{L}}Y_{L}^{\mathrm{in}},\sqrt{2\kappa_{R}}X_{R}^{\mathrm{in}},\sqrt{2\kappa_{R}}Y_{R}^{\mathrm{in}}$),
we write the time-dependent inhomogeneous equations of motion as $\overset{.}{u}(t)=A(t)u(t)+n(t)$,
where $A(t)=$
\begin{equation}
\left[\begin{array}{cccccc}
0 & 1/m & 0 & 0 & 0 & 0\\
-m\Omega_{M}^{2} & -\Gamma_{M} & -G_{xL}(t) & -G_{yL}(t) & G_{xR}(t) & G_{yR}(t)\\
G_{yL}(t) & 0 & -\kappa_{L} & \Delta_{L}(t) & 0 & 0\\
-G_{xL}(t) & 0 & -\Delta_{L}(t) & -\kappa_{L} & 0 & 0\\
-G_{yR}(t) & 0 & 0 & 0 & -\kappa_{R} & \Delta_{R}(t)\\
G_{xR}(t) & 0 & 0 & 0 & -\Delta_{R}(t) & -\kappa_{R}
\end{array}\right].
\end{equation}
In the matrix, $G_{x\sigma}(t)=\eta_{\sigma}\left\langle x(t)\right\rangle $
and $G_{y\sigma}(t)=\eta_{\sigma}\left\langle y(t)\right\rangle $
are the real and the imaginary parts of the scaled coupling constants
$G_{\sigma}(t)=\sqrt{2}\langle a_{\sigma}(t)\rangle\eta_{\sigma}$.
Along with the oscillation of the mechanical resonator, the dynamic
detunings of the two cavities are defined as 
\begin{eqnarray}
\Delta_{\sigma}(t) & = & \Delta_{0,\sigma}\pm\eta_{\sigma}\langle q(t)\rangle,
\end{eqnarray}
where the plus (minus) sign corresponds to the left (right) cavity.

When the tripartite system is stable, it reaches a unique steady state,
independently from the initial condition. Then given any arbitrary
steady state, the fluctuations about it is fully characterized by
its $6\times6$ covariance matrix $V$ of the pairwise correlations
among the quadratures, which obeys the equation $\dot{V}(t)=A(t)V(t)+V(t)A^{T}(t)+D$.
The diagonal elements of the $V$ are, in order, auto-correlations
of the quadratures of the resonator, the left, and the right cavity
mode. Hence, $D=\mathrm{diag}(0,\Gamma_{M}(2\bar{n}+1),\kappa_{L},\kappa_{L},\kappa_{R},\kappa_{R})$
is the diagonal matrix for the corresponding damping and leakage rates
responsible for the fluctuations. The multiple fluctuation-dissipation
relations defined in Sec.~\ref{sec:model} are therefore encapsulated
in the relation $\left\langle n_{i}(t)n_{j}(t')+n_{j}(t')n_{i}(t)\right\rangle /2=\delta(t-t')D_{ij}$.
From its evolution equation, the covariance matrix $V$ can be written
as a block-matrix
\begin{equation}
V=\left[\begin{array}{ccc}
V_{M} & V_{ML} & V_{MR}\\
V_{ML}^{T} & V_{L} & V_{LR}\\
V_{MR}^{T} & V_{LR}^{T} & V_{R}
\end{array}\right],\label{eq:matrix_V}
\end{equation}
where each block represents $2\times2$ matrix. The blocks on the
diagonal indicate the variance within each subsystem (the resonator
$M$, the left cavity mode $L$, and the right cavity mode $R$),
while the off-diagonal blocks indicate covariance across different
subsystems, i.e. the correlations between two components that describe
their entanglement property. 

To compute the pairwise entanglements, we reduce the $6\times6$ covariance
matrix $V$ to a $4\times4$ submatrix $V_{S}$. There are three such
cases of the submatrix $V_{S}$: (i) if the indices $i$ and $j$
for the element $V_{ij}$ are confined to the set $\{1,2,3,4\}$,
the submatrix $V_{S}=[V_{ij}]$ is formed by the first four rows and
columns of $V$ and corresponds to the covariance between the resonator
mode and the left cavity mode. Similarly, (ii) if the indices run
over $\{1,2,5,6\}$, $V_{S}$ is the covariance matrix of the resonator
and the right cavity mode. (iii) If the indices run over $\{3,4,5,6\}$,
$V_{S}$ designates the covariance between the two opposite cavity
modes. Summarizing, the submatrix can be written as
\begin{equation}
V_{S}=\left[\begin{array}{cc}
V_{\alpha} & V_{\alpha\beta}\\
V_{\alpha\beta}^{T} & V_{\beta}
\end{array}\right],\label{eq:matrix_subV}
\end{equation}
where $\alpha$, $\beta$, and $\gamma$ index the subsystems $\{M,L,R\}$
in the optomechanical cavity. The entanglement measured by logarithmic
negativity is computed through a process known as symplectic diagonalization
of each submatrix $V_{S}$, where the entanglement properties are
contained in the symplectic eigenvalues of the diagonalized matrix.
If we write the diagonalized matrix as $\mathrm{diag}(v_{-},v_{-},v_{+},v_{+})$,
then the eigenvalues along the diagonal read~\cite{plenio07}
\begin{equation}
v_{\mp}=\sqrt{\frac{1}{2}\left[\Sigma(V_{S})\mp\sqrt{\Sigma(V_{S})^{2}-4\det V_{S}}\right]},
\end{equation}
where $\Sigma(V_{S})=\det(V_{\alpha})+\det(V_{\beta})-2\det(V_{\alpha\beta})$. 

Denoting the state of a bipartite subsystem in the tripartite optomechanical
cavity as $\rho$, the negativity is defined as
\begin{equation}
N(\rho)=\frac{\left\Vert \text{\ensuremath{\rho}}^{T}\right\Vert _{1}-1}{2},\label{eq:negativity}
\end{equation}
where $\parallel\rho^{T}\parallel_{1}$ indicates the trace norm of
the partial transposition of $\rho$~\cite{Vidal02}. Taking $v_{-}$
as the minimum symplectic eigenvalue of the covariance matrix, $\left\Vert \text{\ensuremath{\rho}}^{T}\right\Vert _{1}$
is equivalent to $1/v_{-}$ after the diagonalization. Hence, the
negativity is a decreasing function of $v_{-}$ and we usually write
$N(\rho)=\max\{0,(1-v_{-})/2v_{-}\}$ and take its logarithmic value
$E_{N}=\ln\left\Vert \text{\ensuremath{\rho}}^{T}\right\Vert _{1}$
as a measure of the entanglement~\cite{Adesso04}. This logarithmic
negativity has the expression $E_{N}=\max\{0,-\ln(v_{-})\}.$

In other words, the symplectic eigenvalue $v_{-}$ completely quantifies
the quantum entanglement between each pair of components in the system.
The necessary condition for showing a bipartite subsystem is entangled
is that the symplectic eigenvalue retains a value less than one, which
is equivalent to the inequality $4\det V_{S}<\Sigma(V_{S})-1/4$~\cite{Simon00}.

\section{Entanglement transfer\label{sec:entanglement_transfer}}

\subsection{Delayed build-up}

\begin{figure}
\includegraphics[bb=0bp 105bp 440bp 465bp,clip,width=8.5cm]{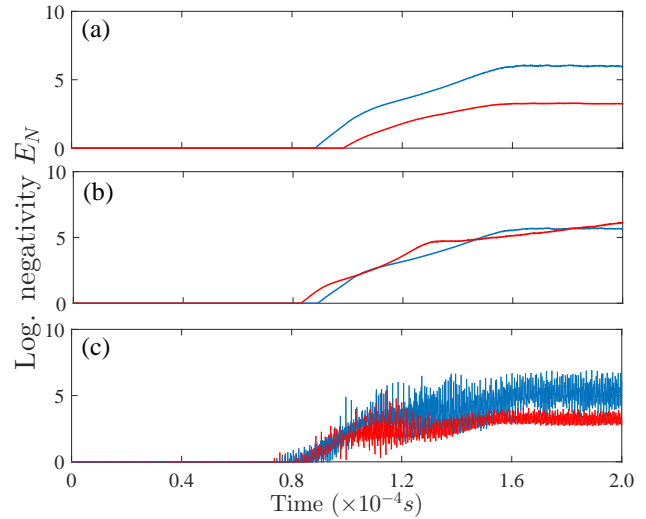}

\protect\caption{(Color online) Time evolution of the tripartite optomechanical system
characterized by the logarithmic negativities between (a) the left
cavity mode and the mechanical resonator, (b) the right cavity mode
and the mechanical resonator, and (c) the left and the right cavity
modes. Two cases are shown with different colors: (i) the blue curves
show the symmetric case where the parameters of the left and the right
cavities are set identical; and, in contrast, (ii) the red curves
show the asymmetric case where some parameters of the two cavities
are set distinct. The parameter values taken for the plots are given
in the text.~\label{fig:comparison}}
\end{figure}

To measure the entanglements, the noise terms $\xi$, $X_{L}^{\mathrm{in}}$,
$Y_{L}^{\mathrm{in}}$, $X_{R}^{\mathrm{in}}$, and $Y_{R}^{\mathrm{in}}$
that appear in the variance matrix of Eq.~(\ref{eq:matrix_V}) are
taken as random variables of zero-mean Gaussian processes. The entanglements
measured in logarithmic negativities $E_{N}$ are plotted against
time for each of submatrices given in Eq.~(\ref{eq:matrix_subV})
to discern the entanglement transfer. We found similar transfer patterns
over a range of parameters close to the experiments\cite{Zhang10}.
One typical case is shown here in Fig.~\ref{fig:comparison}, where
from top to bottom we plot, respectively, $E_{N}$ between the left
cavity and the resonator, between the right cavity and the resonator,
and finally between the left and the right cavities.

For comparison, two cases are plotted for each entanglement pair:
the blue ones denote the symmetric case and the red ones denote the
asymmetric case. For the symmetric case, we adopt for the mechanical
resonator a quality factor $Q=20000$, resonance frequency $\omega_{M}=1$MHZ,
and effective mass $m=10$ ng~; for the cavities, we take cavity
length at $22$ mm with finesse $F=2.6\times10^{5}$ and cavity mode
wavelength of $1064$ nm. We set the power of the driving lasers at
70$\mu\mathrm{W}$, which is detuned from the cavity mode at $\Delta=6.5\omega_{M}$.
For the asymmetric case plotted in red, we have adjusted the right
cavity to a length of $19$ mm, which consequently affects the cavity
leakage and the coupling amplitude between the driving and the cavity,
while the length of the left cavity and other parameters remain unchanged.

We observe from Fig.~\ref{fig:comparison} that there are two phases
in the entanglement evolution. The initial phase is a period of zero
$E_{N}$, showing a delay in the formation of entanglement. The latter
phase is a gradual build-up until certain saturation is reached. While
the entanglement generations between either cavity and the mechanical
resonator are smooth, that between the two cavities are oscillating
or quasi-oscillating because of the nonlinear nature of the radiation
pressure coupling~\cite{wang14}. Averaging out the oscillation,
we see the patterns in the build-up of entanglement are identical
to those between the cavity and the resonator. In addition, the delay
periods among all three pairs coincide, demonstrating the transfer
of cavity-resonator entanglement to intercavity entanglement and showing
that distant entanglement is possible if the distant objects are indirectly
coupled.

The delay in the entanglement build-up, during which $E_{N}$ assumes
zero value, corresponds to the negativity in Eq.~(\ref{eq:negativity})
taking a nonphysical negative value. We can interpret this delay period
as the time duration when the three components in the tripartite system
spend to establish their cooperation, which like the effect of superradiance
depends strongly on the resonance linewidths. Comparing the delays
for the symmetric and the asymmetric cases from Fig.~\ref{fig:comparison}(a)
and (b), we see the similar inverse proportionality in the entanglement
delay $T_{D}$ on the cavity leakage rates $\kappa_{\sigma}$, i.e.,
$T_{D}\propto\kappa_{\sigma}^{-1}$ . When the cavities are setup
symmetrically, we measure the delays in both Fig.~\ref{fig:comparison}(a)
and (b) at about $89\mu\mathrm{s}$; when they are setup asymmetrically
with $\kappa_{L}<\kappa_{R}$, we observe $T_{D}$ for the left cavity
being greater than its counterpart at the right side, at a difference
of $15.7\mu\mathrm{s}$ in time for a difference about $2.3\mathrm{kHz}$
in cavity linewidths.

\subsection{Death and revival}

\begin{figure}
\includegraphics[bb=10bp 110bp 440bp 460bp,clip,width=8.5cm]{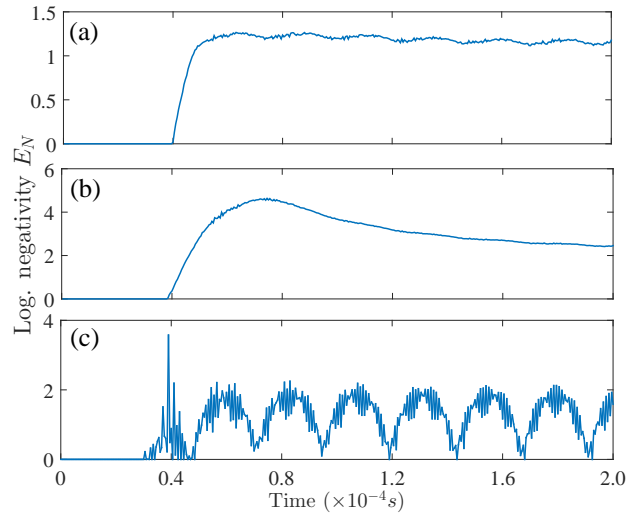}

\protect\caption{Time evolution of the logarithmic negativity $E_{N}$ for the same
three pairs of components in the tripartite system as in Fig.~\ref{fig:comparison}
plotted respectively in (a), (b), and (c), where (c) shows the death
and revival patterns in the intercavity entanglement.~\label{fig:death-revival}}
\end{figure}

The influences of asymmetric parameter setup for the cavities are
not only reflected in the delays of entanglement generation, but also
in the entanglement pattern itself. In Fig.~\ref{fig:death-revival},
we show a typical example with entanglements generated in a pattern
distinctly differently from those in Fig.~\ref{fig:comparison}.
The entanglements measured in logarithmic negativity are again plotted
from top to bottom, respectively, for the three component pairs discussed
above, but with driving laser powers increased to $80\mu\mathrm{W}$
and cavity finesses decreased to $F=1.0\times10^{5}$. The left and
the right cavity lengths remain in an asymmetric setup of $22$mm
and $20$mm, respectively, and the rest of parameters are kept identical
to those in Sec.~\ref{sec:entanglement_transfer}A.

While the cavity-resonator entanglements for the two cavities follow
the pattern of build-up to saturation after a time delay, which is
similar to those of Fig.~\ref{fig:comparison}, the intercavity entanglement
does not but otherwise oscillate over a death-revival cycle. Because
of the inverse proportionality of the time delay to the cavity linewidths,
the plots show a shortened delay and a reduced discrepancy between
the delays in the left and the right cavity-resonator entanglements
due to the decrease in cavity finesses.

On closer inspection, we can see the build-up in (a) and (b) are sharper
and less gradual than their counterparts in Sec.~\ref{sec:entanglement_transfer}A
and the absolute negativity they can obtain are much smaller, especially
for the left cavity. Even for the right cavity, its entanglement with
the resonator declines shortly after a peak value, making all three
plots assume essentially different characteristics than those of Fig.~\ref{fig:comparison}.
This distinction can be attributed to the strong dependence of the
operating regimes of optomechanical systems on external driving power
and cavity finesse. In a single optomechanical cavity, it is reflected
as periodic and quasiperiodic motions of the resonator~\cite{wang14};
in the double optomechanical cavity here, it is reflected as the resonator
being driven monotonically in-phase (Fig.~\ref{fig:comparison}(c))
and driven periodically in-phase and out-of-phase (Fig.~\ref{fig:death-revival}(c))
with the left and right cavities.

\section{Conclusions\label{sec:conclusions}}

To summarize, we have studied the dynamic transfer of quantum entanglement
from those within two cavity-resonator pairs to that between these
two cavities inside a double-cavity optomechanical system. We numerically
solved a coupled set of Heisenberg-Langevin equations to show the
generation of quantum entanglements between each pair of the components
under an experimentally accessible set of parameters. We find that
the entanglement of the indirectly coupled cavities is built up over
time in a pattern similar to those of the directly entangled cavity-resonator
pairs, verifying the entanglement transfer. The similarities are accentuated
by the almost identical characteristic delays and rising patterns
but the entanglement transfer would be suppressed by the asymmetries
in the two cavities. The asymmetries also differentiates the initiation
times of the cavity-resonator entanglements, which leads to our speculation
that the tripartite system is undergoing a cooperation process similar
to that of superradiance before the emergence of the entanglement.
To understand such a transient effect in a multipartite system demands
a detailed analysis of the Heisenberg-Langevin equation set, which
we shall leave to future studies, but we have seen here that dynamic
entanglement is not only a measure of quantum information, but also
a useful tool to dissect the cooperative motions of microscopic systems.
\begin{acknowledgments}
R.~Z. is supported by the National Natural Science Foundation of
China under Grant No.~61463016 and 61340029, Program for New Century
Excellent Talents in University under Grant No.~NCET-13-0795, Landing
project of science and technique of colleges and universities of Jiangxi
Province under Grant No. KJLD14037, Project of International Cooperation
and Exchanges of Jiangxi Province under Grant No.~20141BDH80007.
H.~I. is supported by the FDCT of Macau under grant 013/2013/A1,
University of Macau under grants MRG022/IH/2013/FST and MYRG2014-00052-FST,
and National Natural Science Foundation of China under Grant No.~11404415.
\end{acknowledgments}

\appendix

\section{Steady states of symmetrical double cavity optomechanical system}

Substituting Eq.~(\ref{eq:steady_a}) into Eq.~(\ref{eq:steady_q})
and cancelling the factor $\left\langle q\right\rangle $ on both
sides of the equation, which implies the trivial solution being one
of the steady state in the symmetrical cavity setup, we arrive at
the quartic equation 
\begin{equation}
\left\langle q\right\rangle ^{4}+2\frac{\kappa^{2}-\Delta_{0}^{2}}{\eta^{2}}\left\langle q\right\rangle ^{2}+\left(\frac{\kappa^{2}+\Delta_{0}^{2}}{\eta^{2}}\right)^{2}-\frac{4\Delta_{0}\varepsilon^{2}}{m\eta^{2}\Omega_{M}^{2}}=0.\label{eq:quartic_eq}
\end{equation}
Lacking the odd-order terms in $\left\langle q\right\rangle $, the
roots $\left\langle q\right\rangle ^{2}$ of the equation can be solved
directly through quadratic formula. Since $\kappa^{2}+\Delta_{0}^{2}>0$,
the real roots $\left\langle q\right\rangle $ exist only when:

i) $\left\langle q\right\rangle ^{2}$ is real, i.e. the discriminant
being non-negative, which gives
\begin{equation}
(\eta\varepsilon)^{2}>m\Omega_{M}^{2}\kappa^{2}\Delta_{0},\label{eq:cond_i}
\end{equation}
and ii) the quadratic root $\left\langle q\right\rangle ^{2}$ to
Eq.~(\ref{eq:quartic_eq}) is non-negative.

To satisfy the latter, we have to consider two cases:

ii-1) when $\kappa^{2}-\Delta_{0}^{2}<0$, the square root of the
determinant could take either the positive or the negative value.
For the negative case, it is required that 
\begin{equation}
(\eta\varepsilon)^{2}<m\Omega_{M}^{2}\frac{(\kappa^{2}+\Delta_{0}^{2})^{2}}{4\Delta_{0}}\label{eq:cond_ii_x}
\end{equation}
or ii-2) for the postive case or when $\kappa^{2}-\Delta_{0}^{2}>0$,
it is required that
\begin{equation}
(\eta\varepsilon)^{2}\geq m\Omega_{M}^{2}\frac{(\kappa^{2}+\Delta_{0}^{2})^{2}}{4\Delta_{0}}.\label{eq:cond_ii}
\end{equation}

When two cases of condition (ii) are combined with condition (i),
we see case (ii-1) impose a very stringent constraint on the admissible
values of $(\eta\varepsilon)^{2}$: between zero and $m\Omega_{M}^{2}(\kappa^{2}-\Delta_{0}^{2})^{2}/4\Delta_{0}$.
Case (ii-2) is more inclusive, which is what we are interested in
here. Since it always holds that $(\kappa^{2}+\Delta_{0}^{2})>4\kappa^{2}\Delta_{0}^{2}$,
when the inequality of Eq.~(\ref{eq:cond_ii}) holds, the first condition
in Eq.~(\ref{eq:cond_i}) is automatically satisfied.

To simplify the study, we confine our investigation in the positive
domain of the detuning $\Delta_{0}$, for which Eq.~(\ref{eq:cond_ii})
can be further reduced to
\begin{equation}
\eta\varepsilon\geq\sqrt{\frac{m\Omega_{M}^{2}}{4\Delta_{0}}}(\kappa^{2}+\Delta_{0}^{2}).
\end{equation}

\end{document}